\documentclass[sigconf]{acmart}

\usepackage{array}
\usepackage{multirow}
\usepackage{listings}
\usepackage{algorithm}
\usepackage[noend]{algpseudocode}

\AtBeginDocument{%
  }

\copyrightyear{2025}
\acmYear{2025}
\setcopyright{cc}
\setcopyright{cc}
\setcctype{by}
\acmConference[ISCA '25]{Proceedings of the 52nd Annual International Symposium on Computer Architecture}{June 21--25, 2025}{Tokyo, Japan}
\acmBooktitle{Proceedings of the 52nd Annual International Symposium on
Computer Architecture (ISCA '25), June 21--25, 2025, Tokyo, Japan}\acmDOI{10.1145/3695053.3731095}
\acmISBN{979-8-4007-1261-6/2025/06}

\begin{document}
\title{HPVM-HDC: A Heterogeneous Programming System for Accelerating Hyperdimensional Computing}

\author{Russel Arbore}
\authornote{Equally contributing authors.}
\email{rarbore2@illinois.edu}
\affiliation{
  \institution{University of Illinois Urbana-Champaign}
  \country{USA}
}
\author{Xavier Routh}
\authornotemark[1]
\email{xrouth2@illinois.edu}
\affiliation{
  \institution{University of Illinois Urbana-Champaign}
  \country{USA}
}
\author{Abdul Rafae Noor}
\email{arnoor2@illinois.edu}
\affiliation{
  \institution{University of Illinois Urbana-Champaign}
  \country{USA}
}
\author{Akash Kothari}
\email{akashk4@illinois.edu}
\affiliation{
  \institution{University of Illinois Urbana-Champaign}
  \country{USA}
}
\author{Haichao Yang}
\email{hcyang@ucsd.edu}
\affiliation{
  \institution{University of California San Diego}
  \country{USA}
}
\author{Weihong Xu}
\email{wexu@ucsd.edu}
\affiliation{
  \institution{University of California San Diego}
  \country{USA}
}
\author{Sumukh Pinge}
\email{spinge@ucsd.edu}
\affiliation{
  \institution{University of California San Diego}
  \country{USA}
}
\author{Minxuan Zhou}
\email{mzhou26@iit.edu}
\affiliation{
  \institution{Illinois Institute of Technology}
  \country{USA}
}
\author{Tajana Rosing}
\email{tajana@ucsd.edu}
\affiliation{
  \institution{University of California San Diego}
  \country{USA}
}
\author{Vikram Adve}
\email{vadve@illinois.edu}
\affiliation{
  \institution{University of Illinois Urbana-Champaign}
  \country{USA}
}

\renewcommand{\shortauthors}{Arbore et al.}

\begin{abstract}
%Hyperdimensional Computing (HDC), a technique inspired by cognitive models of computation, has been proposed as an efficient and robust alternative basis for machine learning. The highly parallel nature of HDC algorithms makes them well-suited for execution on several hardware architectures, including CPUs, GPUs, and domain specific accelerators. Retargetable programming systems are crucial for accelerator development, as they allow for parallel development of domain specific algorithms, programs, and accelerator systems. However, HDC programs are often manually written in target-specific languages---no previous programming system enables productive development of HDC programs and generates efficient code for several hardware targets.
Hyperdimensional Computing (HDC), a technique inspired by cognitive models of computation, has been proposed as an efficient and robust alternative basis for machine learning. HDC programs are often manually written in low-level and target specific languages targeting CPUs, GPUs, and FPGAs---these codes cannot be easily retargeted onto HDC-specific accelerators. No previous programming system enables productive development of HDC programs and generates efficient code for several hardware targets.

We propose a heterogeneous programming system for HDC: a novel programming language, HDC++, for writing applications using a unified programming model, including HDC-specific primitives to improve programmability, and a heterogeneous compiler, HPVM-HDC, that provides an intermediate representation for compiling HDC programs to many hardware targets. We implement two tuning optimizations, automatic binarization and reduction perforation, that exploit the error resilient nature of HDC. Our evaluation shows that HPVM-HDC generates performance-competitive code for CPUs and GPUs, achieving a geomean speed-up of 1.17x over optimized baseline CUDA implementations with a geomean reduction in total lines of code of 1.6x across CPUs and GPUs. Additionally, HPVM-HDC targets an HDC Digital ASIC and an HDC ReRAM accelerator simulator, enabling the first execution of HDC applications on these devices.
\end{abstract}



%
% Keywords. The author(s) should pick words that accurately describe
% the work being presented. Separate the keywords with commas.


\maketitle

\section{Introduction}

As machine learning models are hitting the memory and compute limits of modern hardware, they are suffering from low performance and high energy consumption on edge devices. In order to meet the performance, energy, and accuracy demands of modern workloads, a brain-inspired computing paradigm, \textit{Hyperdimensional Computing (HDC)} \cite{kanerva2009hyperdimensional}, is gaining traction. It represents data using \textit{hypervectors}, which are high dimensional vectors with thousands of elements. HDC operations are light-weight, highly parallelizable, and robust to noise, so they are well-suited for computations in resource-constrained environments. As a result, HDC has been used in a plethora of applications including natural language processing \cite{rahimi2016robust}, robotics \cite{mitrokhin2019learning}, voice and gesture recognition \cite{imani2017voicehd, rahimi2016hyperdimensional}, emotion recognition \cite{chang2019hyperdimensional}, graph and hypergraph learning \cite{nunes2022graphhd, hyghd}, DNA sequencing \cite{kim2020geniehd}, recommender systems \cite{guo2021hyperrec}, and bio-signal processing \cite{asgarinejad2020detection}.

HDC algorithms can run on multi-core CPUs \cite{morris2019comphd, verges2023hdcc} and GPUs \cite{kang2022openhd, simon2022hdtorch, kang2022xcelhd} due to their highly parallel character. In recent years, FPGA-based, ASIC-based, and processing-in-memory (PIM) based HDC accelerators have achieved higher performance and energy efficiency \cite{yu2023fully, datta2019programmable, imani2019binary, gupta2018felix, imani2021revisiting, imani2019fach, salamat2020accelerating, tinyHD, DigitalASIC, HDnn-PIM, ReRAMAcc, SpecPCM, rapidoms, spectraflux, msmlcreram}. These accelerators have specialized on-chip buffers and pipelined components to maximize parallelism and minimize data movement. Some devices based on resistive RAM (ReRAM)~\cite{ReRAMAcc} and phase change memory (PCM)~\cite{SpecPCM} also require hardware-specific optimizations to manage trade-offs between accuracy and energy efficiency / performance.

Custom accelerators will only succeed if they are supported by practical, widely adopted programming models and development tools. In the case of HDC, realistic applications must be able to run on custom accelerators as well as more widely available devices, like CPUs and GPUs, because suitable accelerators are not universally available. Even when an accelerator is available, in our experience, accelerators only support a proper subset of core HDC operations, forcing some application components to run on other hardware; moreover, this partitioning between accelerator and general purpose hardware is different for different accelerators.  To achieve this kind of portability and flexible application partitioning, HDC applications require appropriate high-level abstractions to effectively target heterogeneous systems. Without such abstractions, application developers must write their applications to run well on each specific target because an application parallelized for GPU cannot be recompiled and run for an HDC ASIC or for a PIM-based device. Writing such multi-version code would be impractical for production software.
%In contrast, well-designed high-level abstractions for HDC would enable application developers to sufficiently describe their workloads using algorithmic primitives which are easier to map to coarse-grained HDC accelerators and to fine-grained parallelism in CPUs and GPUs, leading to hardware portability. 

There are at least three challenges in designing and supporting high-level, easily portable programming abstractions for HDC applications. 
First, HDC programming requires novel and unusual data encoding schemes, combined with hypervector and hypermatrix operations that depend on those encoding schemes~\cite{heddes2022torchhd}. These custom schemes must be captured in the high-level programming model without putting undue burden on the programmer to implement the encoding-specific details of common operations.
Second, to achieve high performance on many targets, including CPUs, GPUs, and others, programs must use hardware-specific optimizations that maximize data locality and leverage different forms of parallelism in HDC algorithms.
High-level abstractions cannot expose these kinds of optimization options or hardware primitives.
Instead, compilers must automatically apply these optimizations with minimal programmer guidance.
Third, achieving adequate accuracy with HDC algorithms is difficult and hardware-dependent: it requires exploring different algorithmic choices, including the encoding strategy (data-dependent encoding, random projection, custom implementation, etc.), the similarity search schemes (cosine similarity, Hamming distance, etc.), the number of iterations, hypervector length and element precision, and different approximation choices. The optimal choices may vary or be restricted on certain HDC accelerators, forcing difficult hardware-specific tuning. Again, compilers must automatically perform such tuning with limited programmer guidance.

Prior works in HDC algorithm development use general purpose languages such as Python \cite{van1995python, tuomanen2018hands, simon2022hdtorch, heddes2022torchhd, kang2022openhd} and MATLAB \cite{matlab2012matlab, fatica2007accelerating}, and rely on target specific languages for programming accelerators, such as CUDA for GPUs \cite{sanders2010cuda, kim2020geniehd}, and C++ with hardware-specific primitives for FPGAs \cite{salamat2019f5, imani2019fach}. HDCC \cite{verges2023hdcc} is a HDC-specific compiler that can compile high-level descriptions of HDC algorithms to multi-threaded C code with vector extensions, but it's programming model is not expressive and it only targets multi-core CPUs. 
%None of the existing methods for developing HDC applications are inherently retargetable. 
Applications targeting each of the devices previously mentioned are often split into entirely separate programs, possibly in different languages. Programmers must implement and hand-optimize different versions of HDC algorithms using different languages to target various devices, 
%leading to redundancy in engineering time and effort and 
significantly hampering programmer productivity. Moreover, targeting HDC accelerators that support specialized coarse-grained instructions requires writing target-specific code that is very different from the lower-level operations used on CPUs and GPUs.

To address the challenges described above, we propose a C++-based programming language, HDC++ (Section~\ref{sec:hdcpp}), that provides three major benefits. First, it creates a set of common high-level abstractions specifically for HDC that enable programmers to implement HDC algorithms without worrying about low-level target-specific optimizations. Second, it strikes a balance between the complexity of implementing HDC algorithms and flexibility in exploring different algorithmic and approximation choices by providing high-level abstractions that make tweaking algorithms easier while also allowing programmers to implement custom algorithms for their HDC applications. Third, it improves programmer productivity by abstracting away the hardware executing the application---the provided primitives are hardware-independent, so the same HDC++ application can be compiled and executed on CPUs, GPUs, and multiple different HDC accelerators \textit{without code modification}. We have ported five applications to HDC++.

We also propose HPVM-HDC (Section~\ref{sec:hpvmhdc}), a compiler based on Heterogeneous Parallel Virtual Machine (HPVM)~\cite{kotsifakou2018hpvm, ejjeh2022hpvm} that compiles hardware agnostic HDC++ applications automatically onto CPUs, GPUs, a digital HDC ASIC and a simulated ReRAM HDC accelerator. HDC++ programs are lowered to a retargetable intermediate representation (IR) with HDC-specific operations (Section~\ref{sec:hd-frontend}). The HPVM-HDC IR structurally captures the parallelism in HDC algorithms, enabling compilation onto hardware targets with coarse-grained instructions such as HDC accelerators as well as targets exposing fine-grained parallelism such as GPUs. Additionally, HPVM-HDC IR supports domain-specific optimizations that exploit the inherent trade-off in HDC between end-to-end quality of service and performance. We implement two such optimizations: automatic binarization, which enables programmers to automatically modify hypervector lengths and precisions to use efficient 1-bit level computations, and reduction perforation, which skips elements in hypervectors when computing reductions. Such approximations lead to significant speedups with little programmer effort and marginal loss in application-level quality of service (Section~\ref{sec:hd-opt}). HPVM-HDC lowers HDC-specific operations through target specific backends for a wide variety of hardware targets: CPUs, GPUs, a taped-out digital HDC ASIC, and a simulated Resistive RAM (ReRAM) HDC accelerator (Section~\ref{sec:hd-backend}). Baseline implementations of our evaluated HDC applications exist only for CPUs and GPUs (and are target-specific)---HPVM-HDC achieves competitive performance with those baseline implementations \emph{using a single retargetable implementation}, achieving a geomean speed-up of 1.17x against optimized baseline CUDA implementations on GPUs. For the accelerators we compile applications to using HPVM-HDC, \textit{no prior evaluation has been performed with a full HDC application}.

Overall, this paper makes the following contributions.

\begin{itemize}
    \item The first set of high-level and expressive programming abstractions for writing HDC programs easily without the need to specialize them for specific hardware targets and without losing performance or accuracy (Section~\ref{sec:hdcpp}).
    
    \item The first retargetable compilation framework for HDC, HPVM-HDC, which compiles HDC++ programs and generates high-performance code for a variety of hardware targets including CPUs, GPUs, a digital HDC ASIC, and a simulated ReRAM HDC accelerator (Section~\ref{sec:hpvmhdc}).
    
    \item An evaluation of the programming system on multiple HDC applications (HD-Classification, HD-Clustering, HyperOMS, RelHD, and HD-Hashtable). HPVM-HDC achieves competitive performance with existing target-specific baseline implementations, achieving a geomean speed-up of 1.17x against optimized baseline CUDA codes on GPUs. Additionally, we offer the first demonstration of executing certain applications on recent HDC accelerators for which there are no baseline implementations (Section~\ref{sec:eval-perf}).
    
    \item An evaluation of domain specific approximation-based compiler optimizations with HPVM-HDC that are supported on the CPU and GPU targets. We observe up to 3.4x performance improvement of HDC inference on the GPU with marginal loss in end-to-end accuracy (Section~\ref{sec:eval-opt}).
    
    \item A thorough analysis of programmability with HDC++. We observe a 1.6x reduction in the total number of lines of codes needed to express HDC applications when using HDC++ over baseline codes. Additionally, common tuning edits take seconds in HDC++, compared to minutes or hours in baseline codes. (Section~\ref{sec:eval-programmability}).
\end{itemize}

\section{Background}

\subsection{Hyperdimensional Computing}
\label{sec:hdc_explain}

Hyperdimensional Computing (HDC) is a machine learning paradigm inspired by cognitive models of computation. HDC algorithms operate by encoding input features into a high dimensional representation called hypervectors. In contrast to Deep Neural Networks (DNNs) which rely on expensive feed-forward computations for inference, HDC relies on encoding query vectors into a high dimensional space and searching through a class hypervector database representing the model---the most similar class hypervector signifies the predicted label. Both of these operations are extremely lightweight in comparison to repeated matrix multiplications. Additionally, HDC training uses repeated element-wise addition instead of back-propagation.

For several tasks, such as graph and hypergraph learning \cite{nunes2022graphhd, hyghd} and federated learning \cite{federated}, HDC techniques have been proposed that achieve significant improvements in latency and energy efficiency over prior DNN-based approaches while maintaining competitive quality of service. While these improvements are significant when deployed on GPU-based systems, HDC solutions have been shown to achieve orders of magnitude improvements in power usage when deployed on specialized hardware \cite{hyghd, DigitalASIC, ReRAMAcc}. Additionally, prior work combines HDC techniques with DNN approaches to achieve both high accuracy and energy efficiency \cite{ReRAMAcc, federated}.

Fine-tuning HDC hyper-parameters (such as element representation, encoding dimensionality and training iterations) and algorithm choices play a much more important role in practice for achieving good quality with HDC in comparison to DNNs, where back-propagation can compensate for sub-optimal choices during training. HDC algorithms generally vary in the used encoding schemes, search algorithms, and training procedures which are often application specific. These different algorithms can be used to implement many kinds of applications, including classification \cite{HD2FPGA}, clustering \cite{hdclustering}, and GNN learning \cite{relhd} (see Table~\ref{tab:HDC-workloads}). The lightweight and parallel nature of HDC enables high performance across hardware targets. 

%Manually implementing HDC algorithms involves numerous challenges. Managing high-dimensional hypervectors (1-10k dimensions), including their generation and storage, as well as performing key HDC operations such as element-wise addition, multiplication, permutation, and similarity comparison, is computationally demanding and often memory-bound. 
Optimizing HDC operations for specific hardware targets and ensuring parallelization, efficient data reuse, and maintaining cache hierarchies requires low-level adjustments that are difficult to generalize across platforms, especially without domain and hardware expertise. Similarly, choosing the best encoding scheme for an application requires domain expertise.
%The probabilistic nature of HDC further complicates debugging and verification, making reliability hard to achieve without automated tools.

Adapting HDC applications to new frameworks and accelerators often necessitates redefining encoding schemes and retraining models, which can be both time-consuming and suboptimal. For instance, HyperSpec \cite{hyperspec} and HyperOMS \cite{hyperoms} are two large-scale data analytics workloads in the domain of proteomics, specifically for proteomics data clustering and library searching. The scalability of these two applications are limited due to repeated data movement. Prior works~\cite{spechd, rapidoms} have used near-storage FPGA implementations to address these challenges, however require extensive manual effort in rewriting and verification,  optimization of memory access patterns, low-level hardware interfacing, and ensuring computational correctness of algorithms under constrained resources. %To address this, two works utilizing near-storage FPGA implementations using HLS were proposed \cite{spechd, rapidoms}. The approach presented in these works requires extensive manual effort in rewriting and verification, involving complex tasks such as optimization of memory access patterns, low-level hardware interfacing, and ensuring computational correctness of algorithms under constrained resources. Prior works have lacked a programming system specifically for HDC that would enable efficient, scalable, and reliable implementations of HDC applications while abstracting hardware-specific details and providing a robust foundation for future architectural and application development efforts.

\begin{figure}
\centering
\begin{minipage}{.25\textwidth}
  \centering
  \includegraphics[width=0.8\linewidth,height=5.5cm]{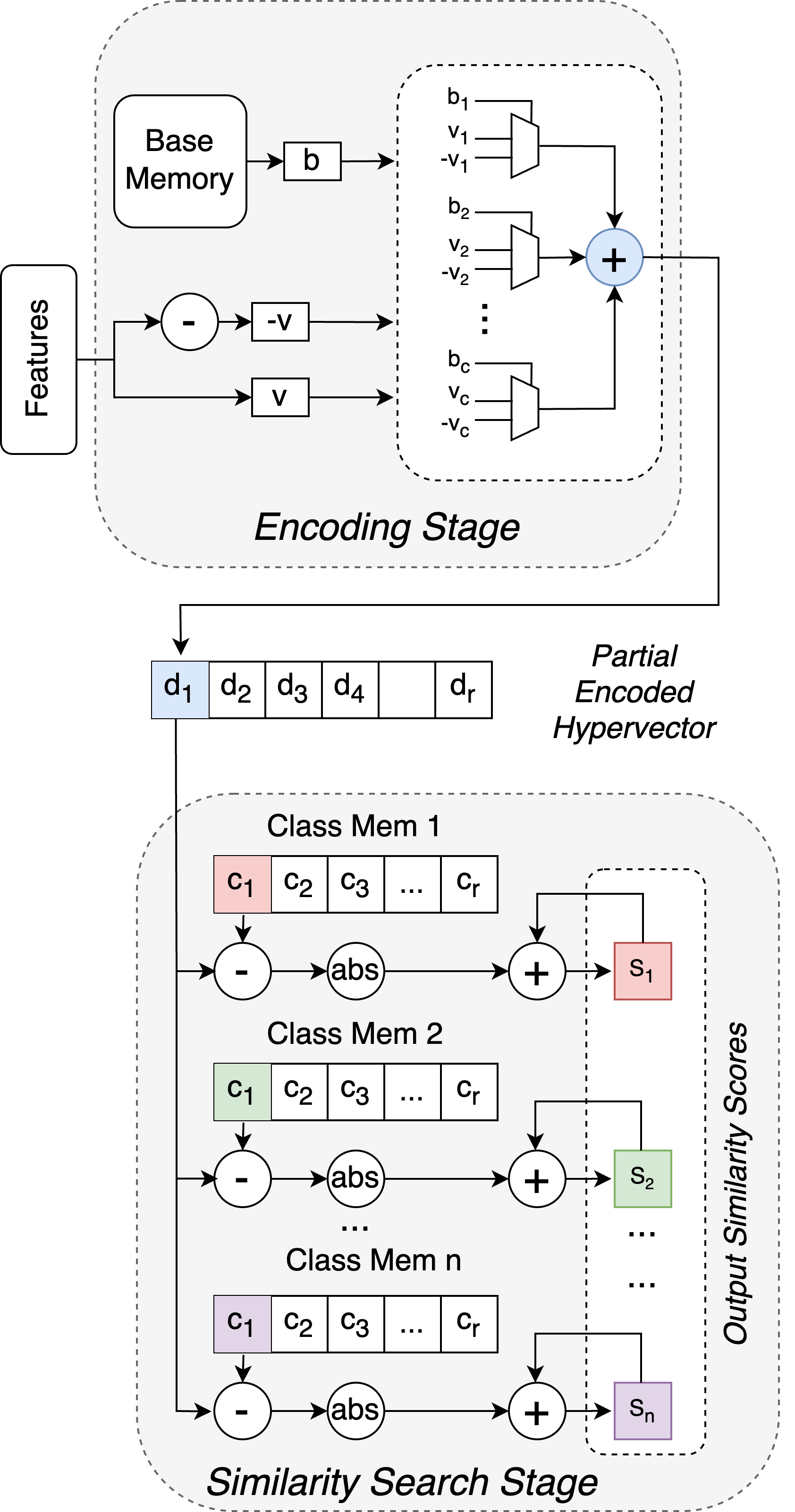}
\end{minipage}%
\begin{minipage}{.25\textwidth}
  \centering
  \includegraphics[width=1.0\linewidth]{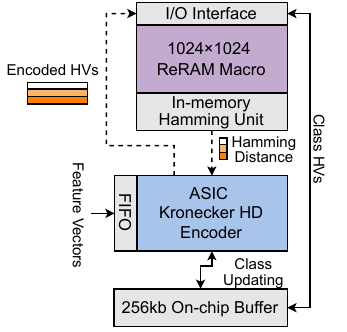}
  
  \label{fig:test2}
\end{minipage}
\caption{Architecture block diagrams for the digital HDC ASIC (left) and ReRAM HDC accelerator (right).}
\label{fig:tiny-hd}
\end{figure}

\subsection{HDC Hardware Accelerators}

HPVM-HDC compiles to two custom HDC accelerators. In this section we briefly describe the devices. Figure~\ref{fig:tiny-hd} shows diagrams for their architectures.

\textbf{\textit{HD Digital ASIC}}: We run HDC workloads on an ASIC chip fabricated in a 40nm technology. The architecture of the ASIC is described by Yang et al. \cite{DigitalASIC}. The device supports cyclic random projecting operations for encoding and pipelined Hamming distance for both training and inference. In this work, we just use the HDC module, which achieves 0.78 TOPS/W (6.6x higher than prior state-of-the-art). The ASIC provides a functional interface for device configuration, data movement, training, and inference. An FPGA is directly wired to the ASIC and an ARM CPU to facilitate communication---the ARM CPU acts as the host and the ASIC has its own private memory. See Listing~\ref{lst:acc_code} for example code generated by HPVM-HDC targeting this interface. Execution times are measured on the chip.

\textbf{\textit{HD ReRAM}}: We also run HDC programs on a resistive RAM (ReRAM) based HDC accelerator. The architecture of the accelerator is described by Xu et al. \cite{ReRAMAcc}, and has a similar programming interface to the digital ASIC. The device supports "tensorized" encoding, which is a more energy efficient variant of random projection encoding, summation based one-shot training, and Hamming distance for inference. Hamming distances are progressively computed until the relative ranking between candidate hypervectors can no longer change. The encoding and inference stages are accelerated using a large ReRAM array. We used a simulator of the device to emulate the performance of HDC code executing on ReRAM hardware. The simulator uses extracted timing and energy parameters from commercial SRAM and ReRAM macros in the 40nm technology node provided by a foundry.

Although these accelerators use different processing technologies to accelerate HDC workloads, their programming interfaces are similar. They expose high-level operations to the host machine, such as run one iteration of training given a single data point or infer the label for a single hypervector given pre-programmed class hypervectors. Other proposed HDC accelerators \cite{SpecPCM} expose a similar programming model. These instructions are difficult to generate from existing HDC codes. Monolithic encoding, inference, and training instructions are too high-level to automatically identify in low-level application code commonly found in prior work on HDC algorithms, as compilers cannot typically reason with enough certainty about the algorithms a user may implement.

\begin{figure}
    \centering
    \includegraphics[width=0.5\textwidth,height=5cm]{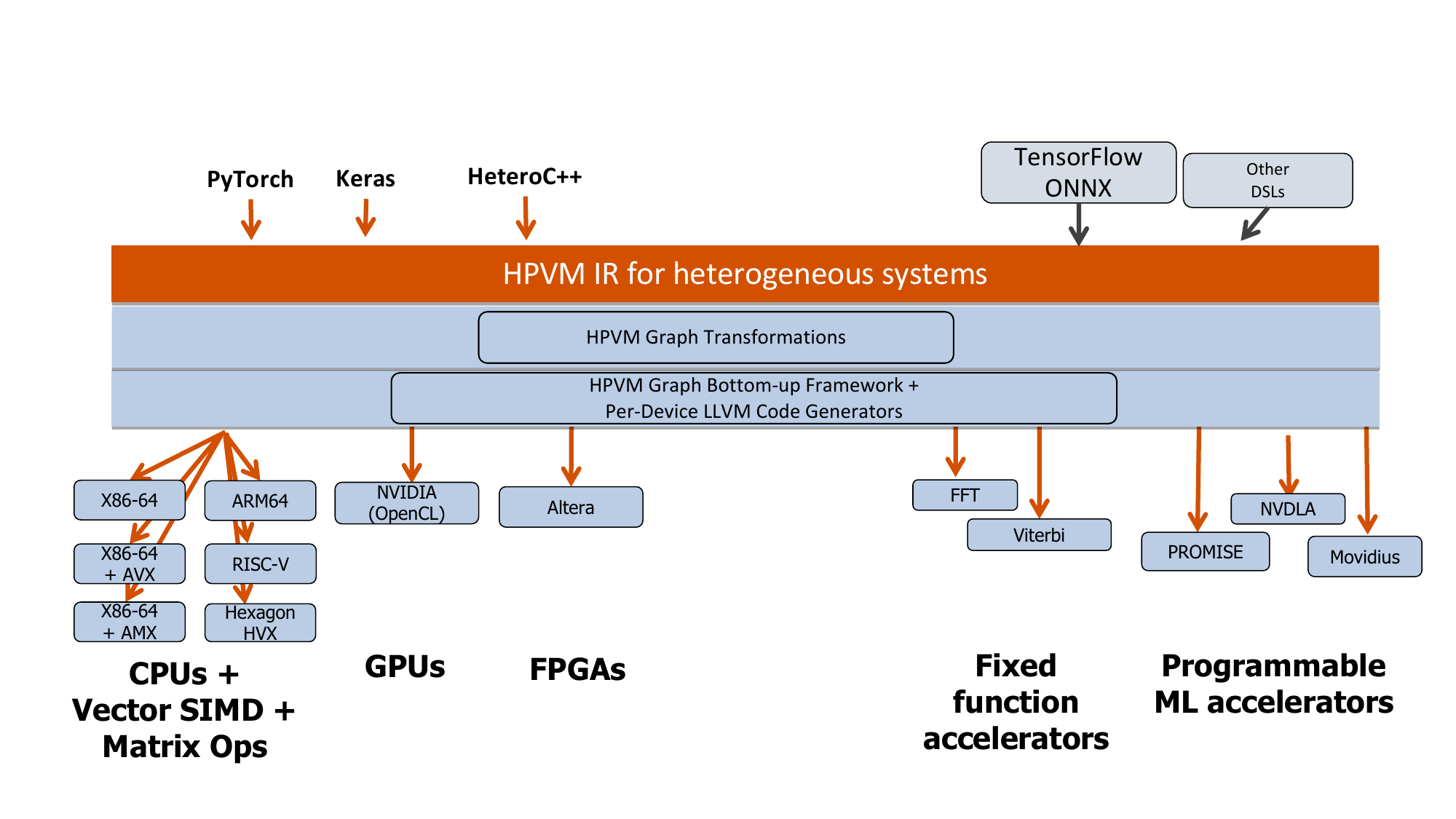}
    \vspace{-.1in}
    \caption{The HPVM Compiler Infrastructure \cite{ejjeh2022hpvm}.}
    \label{fig:hpvm-overview}
\end{figure}

\begin{figure}
    \centering
    \includegraphics[width=0.46\textwidth,height=4.3cm]{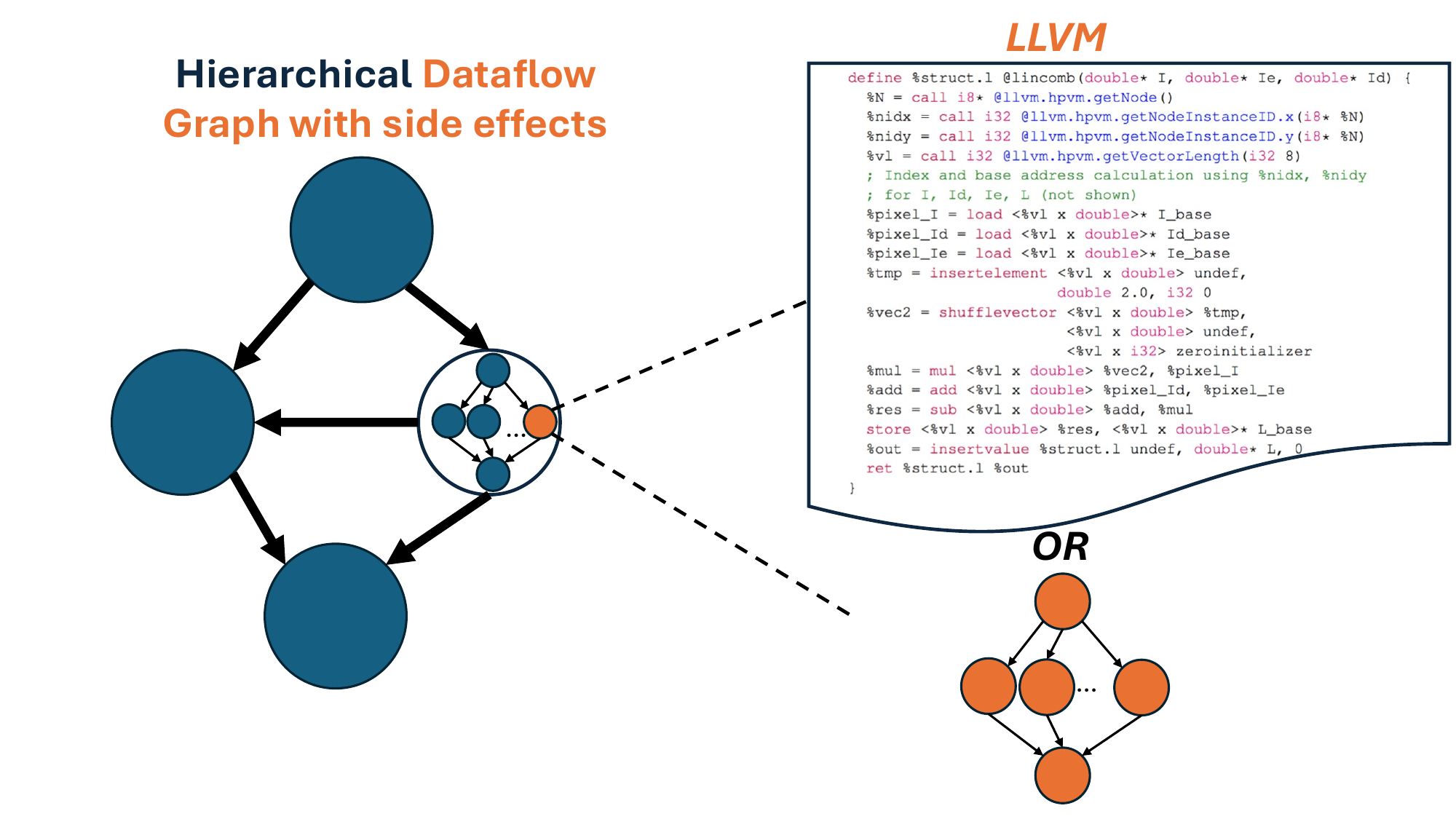}
    \caption{Hierarchical representation of code in HPVM. Each node in the dataflow graph corresponds to either a leaf node performing compute (LLVM function) or a nested sub-graph. Edges between nodes correspond to logical data transfers.}
    \label{fig:hpvm-ir}
\end{figure}

\subsection{HPVM Compiler Infrastructure}
The HPVM Compiler Infrastructure \cite{ejjeh2022hpvm} extends the LLVM compiler infrastructure with a retargetable parallel program representation for heterogeneous architectures (as shown in Figure~\ref{fig:hpvm-overview}). HPVM IR provides constructs for representing various forms of parallelism including task-level, data-level, and streaming parallelism (and expresses vector parallelism in the LLVM code). Parallel programs are represented using a hierarchical directed acyclic graph (DAG). Nodes in the graph fall into two categories: internal nodes, which capture hierarchical parallelism by containing entire sub-graphs, and leaf nodes, which represent individual units of computation, represented as LLVM IR (shown in Figure~\ref{fig:hpvm-ir}). Edges between nodes in the HPVM DAG represent \emph{logical} data transfers, i.e., explicit data copies may or may not be needed. Additionally, each node is annotated with one or more hardware targets, and code is generated by the appropriate back ends for each of those targets (possibly multiple targets).

\subsection{Hetero-C++}
Hetero-C++ is a parallel C++ dialect, with parallel features similar to a subset of OpenMP (primarily for ease of porting C++ applications without a full OpenMP front-end).
Hetero-C++ compiles to HPVM IR \cite{ejjeh2022hpvm}. Hetero-C++ provides a high-level, target-agnostic programming interface for capturing the various forms of parallelism in an application. A function in Hetero-C++ may be split into multiple parallel tasks and parallel loops which are individually lowered to distinct HPVM IR nodes. These distinct components in a single application are individually compiled to different targets as specified by the user, easing programmability for applications on heterogeneous systems and improving portability across systems. Both HDC++ and Hetero-C++ use marker functions instead of pragmas to retain placement within the emitted LLVM IR.

\definecolor{codegreen}{rgb}{0,0.6,0}
\definecolor{codegray}{rgb}{0.5,0.5,0.5}
\definecolor{codepurple}{rgb}{0.58,0,0.82}
\definecolor{backcolour}{rgb}{0.95,0.95,0.92}

\lstdefinestyle{mystyle}{
    backgroundcolor=\color{backcolour},   
    commentstyle=\color{codegreen},
    keywordstyle=\color{magenta},
    numberstyle=\tiny\color{codegray},
    stringstyle=\color{codepurple},
    basicstyle=\ttfamily\footnotesize,
    breakatwhitespace=false,         
    breaklines=true,                 
    captionpos=b,                    
    keepspaces=true,                 
    numbers=left,                    
    numbersep=5pt,                  
    showspaces=false,                
    showstringspaces=false,
    showtabs=false,                  
    tabsize=2
}

\lstset{style=mystyle}

\lstset{emph={%  
    hypervector, hypermatrix%
    },emphstyle={\color{red}\bfseries}%
}%

\begin{table*}[!h]
  \centering
  \resizebox{\linewidth}{!}{%
  \begin{tabular}{|p{0.5\linewidth}|p{0.5\linewidth}|}
    \hline
    \textbf{HDC Algorithmic Primitive} & \textbf{Description}  \\
    \hline
    \texttt{\{HV,HM\} hyper\{vector,matrix\} (void)} &  Initialize an empty hypervector / hypermatrix \\
    \hline
    \texttt{\{HV,HM\} create\_hyper\{vector,matrix\}(Function init)}  &  Initialize a hypervector / hypermatrix with an initialization function \\
    \hline
    \texttt{\{HV,HM\} random\_hyper\{vector,matrix\} (void)} &  Initialize a hypervector / hypermatrix with random values \\
    \hline
    \texttt{\{HV,HM\} gaussian\_hyper\{vector,matrix\}(void)}  &  Initialize a hypervector / hypermatrix with gaussian distribution. \\
    \hline
    \texttt{HV wrap\_shift(HV input, int shift\_amount)}  &  Rotate with wrap-around elements in hypervector \\
    \hline
    \texttt{\{HV,HM\} sign (\{HV,HM\} input)}  &  Maps each element to {+1, -1} by checking the sign of the element. \\
    \hline
    \texttt{\{HV,HM\} sign\_flip (\{HV,HM\} input)}  &  Flips the sign of each element in a hypervector / hypermatrix \\
    \hline
    \texttt{\{HV,HM\} \{add,sub,mul,div\} \{(HV,HV) lhs, (HM,HM) rhs\}} &  Element-wise binary operators between hypervectors / hypermatrices \\
    \hline
    \texttt{\{HV,HM\} absolute\_value  (\{HV,HM\} input)} &  Element-wise absolute value operation on  hypervector / hypermatrix \\
    \hline
    \texttt{\{HV,HM\} cosine (\{HV,HM\} input)} &  Element-wise cosine operation on  hypervector / hypermatrix \\
    \hline
    \texttt{\{T,HV\} l2norm (\{HV,HM\} input) }  &  Calculate the L2 Norm for hypervector / hypermatrix. \\
    \hline
    \texttt{T get\_element (\{HV,HM\}, int row\_idx[, int col\_idx])}  & Indexing into a hypervector / hypermatrix. \\
    \hline
    \texttt{\{HV,HM\} type\_cast (\{HV,HM\} input, T ty)} & Type cast hypervector / hypermatrices elements to result type \\
    \hline
    \texttt{\{T,HV\} arg\_\{min,max\}(\{HV,HM\} input)} & Return the arg max or min of a hypervector / per row of hypermatrix \\
    \hline
    \texttt{void set\_matrix\_row(HM mat, HV new\_row, int row\_idx)}  & Update a specified row in the hypermatrix with a provided hypervector \\
    \hline
    \texttt{HV get\_matrix\_row (HM mat, int row\_idx)}  & Get a specified row in the hypermatrix producing a hypervector \\ 
    \hline
    \texttt{HM matrix\_transpose (HM input)} & Transpose hypermatrix \\
    \hline
    \texttt{\{T,HV,HM\} cossim(\{HV,HM\} lhs, \{HV,HM\} rhs)} & Perform cosine similarity between hypervectors \\
    \hline
    \texttt{\{T,HV,HM\} hamming\_distance(\{HV,HM\} lhs, \{HV,HM\} rhs)} & Perform Hamming distance between hypervectors \\
    \hline
    \texttt{\{HV,HM\} matmul(\{HV,HM\} lhs, HM rhs)} & Matrix multiplication between hypervectors and hypermatrices \\
    \hline
    \texttt{void red\_perf(\{T,HV,HM\} result,int begin, int end, int stride)} & Annotates HDC primitive returning \texttt{result} with description of reduction perforation (Section ~\ref{sec:hd-opt}).  \\
    \hline
    \texttt{HM encoding\_loop(Func encode, HM queries, HM encoder)} & Apply HDC encoding over entire dataset \\
    \hline
    \texttt{int[] inference\_loop(Func infer, HM queries, HM classes)} & Apply HDC inference over entire dataset \\
    \hline
    \texttt{void training\_loop(Func train, HM queries, int[] labels, HM classes, int epochs)} & Apply HDC training over entire dataset \\
    \hline
  \end{tabular}
  }
  \caption{HDC primitives in HDC++. \texttt{HV} and \texttt{HM} denote the hypervector and hypermatrix types respectively. \texttt{T} denotes an element type, which is a signed scalar type (any of \texttt{int8\_t}, \texttt{int16\_t, \texttt{int32\_t}}, \texttt{int64\_t}, \texttt{float}, or \texttt{double}).}
  \label{tab:hdc-primitives}
\end{table*}

\section{The HDC++ Language}
\label{sec:hdcpp}

HDC++ is a domain specific language built on top of Hetero-C++ \cite{ejjeh2022hpvm}. 
Note that the choice of this base language is not essential in our work: any compiled parallel language with explicit task and data parallelism would achieve the same purpose.
HDC++ introduces primitives for creating and manipulating hypervectors and hypermatrices, allowing developers to easily develop HDC algorithms inside their applications. Listing~\ref{lst:simple} shows how HDC++ can be used to express classification inference. First, random projection encoding is used to encode input features into a 2048-dimensional representation. Then, Hamming distance is applied between the encoded hypervector and the class hypervectors (represented using a hypermatrix datatype) to produce a vector of dissimilarities. In this example, the result is a 26 dimensional vector, as there are 26 output label classes. Finally, the HDC \texttt{arg\_min} operation identifies the corresponding class hypervector which is least dissimilar to the encoded hypervector.

\begin{lstlisting}[language=C, label={lst:simple}, caption={HDC++ example for random-projection encoding and dissimilarity between query and class hypervectors.}]
hypervector<617> input_features = ...; 
hypermatrix<2048, 617> rp_matrix = ...;
hypermatrix<26, 2048> clusters_matrix = ...;
hypervector<2048> encoded_data = __hetero_hdc_matmul(input_features, rp_matrix);
hypervector<26> dists = __hetero_hdc_hamming_distance(encoded_data, clusters_matrix);
int label = __hetero_hdc_arg_min(dists);
\end{lstlisting}

In addition to HDC specific primitives, application developers writing HDC++ can use Hetero-C++ primitives for expressing generic parallel code. HDC applications exhibit parallelism both within the HDC algorithmic primitives (operator data-level parallelism), as well as across operations (task-level or outer-loop-level parallelism). By enabling HDC++ to interoperate with Hetero-C++, application developers can compactly describe complex HDC algorithms using HDC primitives while describing inter-operator parallelism using Hetero-C++ and keeping the application retargetable. For example, the HyperOMS~\cite{hyperoms} application uses level ID encoding, which has an outer parallel loop that isn't captured by HDC primitives, but is captured by Hetero-C++'s parallel loop annotation. HDC++ can express all of the components of the HyperOMS application in a parallel fashion, providing both programmability and performance.

HDC++ contains 24 HDC primitives for expressing HDC algorithms in a target-agnostic programming interface. The primitives are parameterized by the element type and the dimensionality of the involved hypervectors and hypermatrices. A comprehensive list of the HD primitives is presented in Table~\ref{tab:hdc-primitives}.

\subsection{High-Level Stage Primitives}

In addition to describing individual HDC operations, HDC++ enables describing three commonly used HDC algorithmic \textit{stages}: \texttt{encoding\-\_loop}, \texttt{training\-\_loop}, and \texttt{inference\-\_loop}. Listing~\ref{lst:stage} illustrates expressing an HDC inference loop using the corresponding primitive. There are two advantages of providing such high-level primitives. First, it enables HPVM-HDC to target HDC accelerators, which provide coarse-grain algorithm acceleration, in contrast to acceleration of finer-grain HDC operations. Second, the use of loop intrinsics for the key algorithmic stages implemented by the HDC accelerators allows HPVM-HDC to elide data movements to and from the accelerators. Crucially, these high-level stage primitives are composable with more granular primitives in HDC++---this allows us to express certain applications that map partially, but not fully to HDC accelerators, such as HD-Clustering. By writing HD-Clustering in HDC++, we are able to map the computationally intensive part of clustering, HDC inference, to HDC accelerators, while performing more ancillary tasks, such as cluster updates and the initial generation of the random projection matrix, on the CPU or GPU.

The algorithmic stage primitives in HDC++ each take as input an "implementation" function. This function implements encoding, training, or inference using more granular HDC primitives, and is executed when targeting CPUs or GPUs, rather than HDC accelerators. This is because while HDC accelerators implement specific encoding, training, and inference algorithms, CPUs and GPUs can be programmed to implement a variety of algorithms---it is up to the application developer to choose concrete versions of these algorithms for these targets.

\begin{lstlisting}[language=C, label={lst:stage}, caption={HDC++ example of an inference loop. \texttt{QueriesHM} are input vectors to classify, \texttt{ClassHM} contains representative hypervectors per class, \texttt{InferenceResults} is the destination pointer to store the results in, and \texttt{InferenceFunction} is a function implementing the inference algorithm using HDC primitives.}]
__hetero_hdc_inference_loop(InferenceFunction, QueriesHM, ClassHM, InferenceResults);
\end{lstlisting}

\section{The HPVM-HDC Compiler}
\label{sec:hpvmhdc}

% Figure Source: https://docs.google.com/presentation/d/1G0_Z5UZ9aQu_EcKXL8XtbaxyGAksf821jCDOpZwlB4Y/edit?usp=sharing
\begin{figure*}
    \centering
    \includegraphics[width=0.63\textwidth]{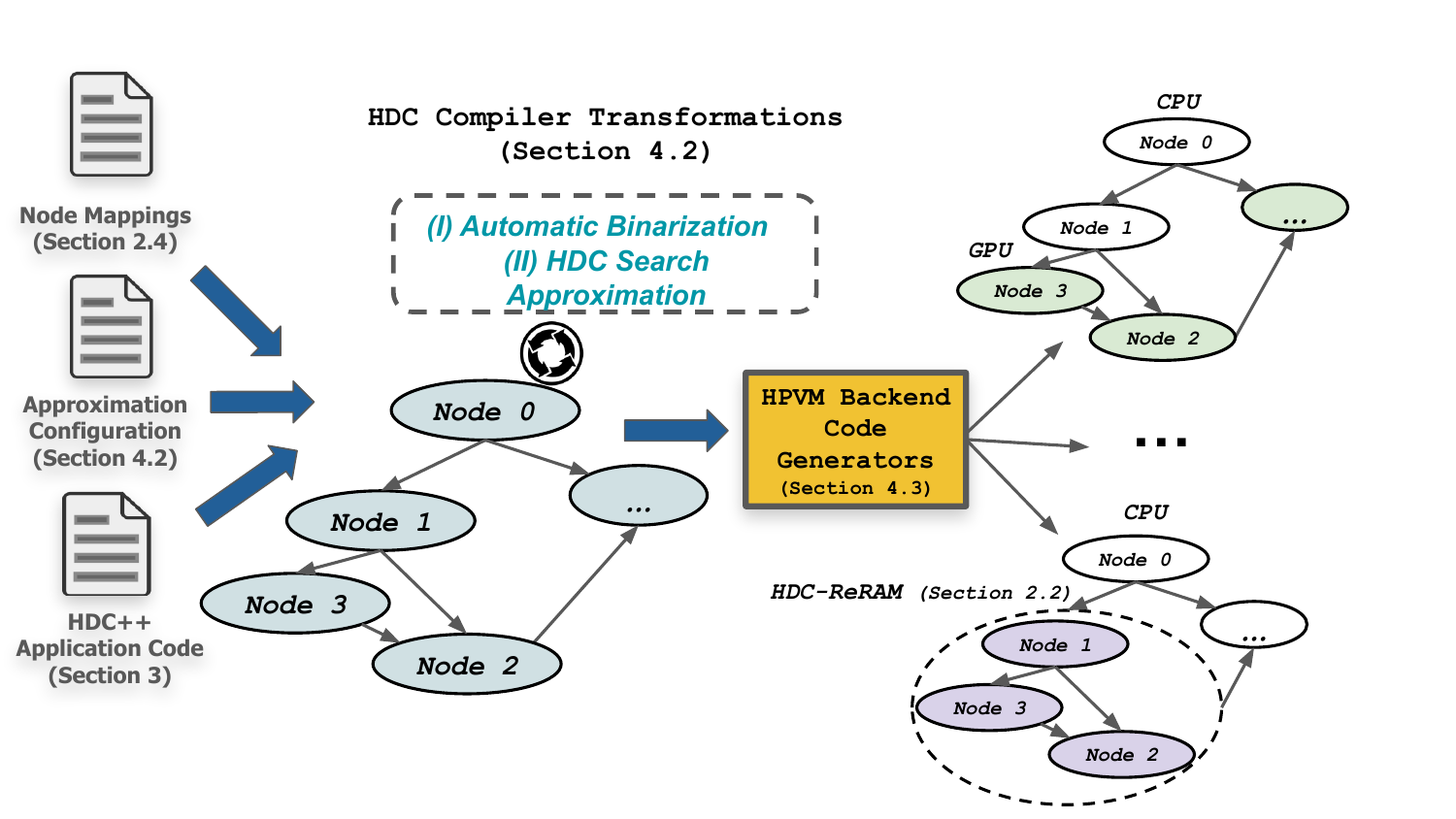}
    \caption{Compilation workflow of an HDC++ application using the HPVM-HDC compiler. First the HDC++ application is compiled to HPVM IR with HDC primitives. Optional HDC optimizations are performed. Finally, the application is compiled to different hardware targets. Different nodes in a DFG may be lowered to different hardware targets.}
    \label{fig:overview}
\end{figure*}

Figure \ref{fig:overview} illustrates the compilation workflow for HDC++ applications onto various target architectures. HPVM-HDC compiles HDC++ application code to an intermediate representation for HDC that is extended from the base HPVM IR (Section ~\ref{sec:hd-frontend}). Then, optimizations are optionally performed (Section ~\ref{sec:hd-opt}). Finally, the program is compiled to the user-specified target, generating the appropriate device and host code (Section ~\ref{sec:hd-backend}).

\subsection{The HPVM-HDC IR}
\label{sec:hd-frontend}

HDC++ application code is compiled into an IR that contains HDC primitives as well as generic parallel task and loop primitives. The HDC primitives are represented as LLVM intrinsic functions which represent the same operations as the HDC++ primitives listed in Table~\ref{tab:hdc-primitives}. We use HPVM IR primitives to represent task and data parallelism. The combination of these two kinds of primitives on top of LLVM IR form HPVM-HDC IR. Lowering HDC++ into HPVM-HDC IR is simple. HPVM already has infrastructure for transforming opaque function calls in Hetero-C++ into LLVM intrinsics; we use this functionality to lower HDC primitives, which are opaque function calls in C++, to LLVM intrinsics. These HDC primitives become intrinsics inside the HPVM IR node functions.

HDC primitive intrinsic functions in HPVM-HDC IR can be lowered in two different ways, depending on the targeted device. They may be lowered by translating intrinsics into HPVM IR subgraphs representing the underlying computation of those HDC operations. As an example, consider compiling a Hamming distance primitive (Listing~\ref{lst:hamming-impl-s1}). HPVM-HDC can lower the primitive into a two deep loop nest, where the outer loop iterates over the classes dimension and the inner loop iterates over the hypervector dimension (Listing~\ref{lst:hamming-impl-s2}). HPVM represents parallel loop iterations as a set of dynamic instances of a node function, each with a unique ID corresponding to the loop index. In this example, the outer loop is known to be parallel, and is outlined into a separate function. The parallel loop iteration is identified by first obtaining the handle to the node in the DFG (line 6) and then querying the dynamic instance ID (line 7). Additional annotations are emitted on line 5 specifying \textit{data} and \textit{classes} as input edges and \textit{result} as an input and output edge. In this scenario, HDC primitives are lowered into generic HPVM IR before the compiler backends are invoked.

\begin{lstlisting}[language=C, label={lst:hamming-impl-s1}, caption={Hamming Distance calculation in HDC++.}]
void get_dissimilarity(hypervector<2048> data,
    hypermatrix<26, 2048> classes,
    hypervector<26> result)
    result= __hetero_hdc_hamming_distance(data, classes);
}
\end{lstlisting}

\begin{lstlisting}[language=C, label={lst:hamming-impl-s2}, caption={Hamming Distance calculation in HPVM IR.}]
void get_dissimilarity_parallel_node(
    hypervector<2048>  data, 
    hypermatrix<26, 2048>  classes, 
    hypervector<26>  result)
    __hpvm__attributes(3,data,classes,result,1,result);
    void* NodeID = __hpvm__getNode();
    int i = __hpvm__getNodeInstanceID_x(NodeID);
    float distance = 0;
    for(int j = 0; j < 2048; j++)
        distance += data[j] == classes[i][j] ? 0 : 1;
    result[i] = distance;
    __hpvm__return(1, result);
}
\end{lstlisting}

Alternatively, HDC primitives may instead be lowered by translating the intrinsics into calls to CUDA kernels, cuBLAS \cite{cublas} calls, Thrust \cite{thrust} calls, or calls to an accelerators' functional interface. These primitives are not lowered into generic HPVM IR. Instead, the compiler backends will generate the appropriate code from the HDC primitives directly. HPVM-HDC decides which lowering strategy to take based on which backend is invoked by the user (see Section~\ref{sec:hd-backend}).

\subsection{HDC Approximation Optimizations}
\label{sec:hd-opt}
The HDC paradigm is robust to error, both in inputs and in processing \cite{kanerva2009hyperdimensional}---this enables various approximation techniques on HDC applications. We implement two such optimizations in HPVM-HDC for transforming HDC++ applications.

\textit{\textbf{Automatic Binarization Propagation}}. The HDC paradigm requires exploring different choices for the size of the encoding dimension, element types, and choice of HDC operations. As HDC++ primitives are parameterized by types, the application developer may choose to represent the resultant hypervector using the same element type as the input features (e.g. float) or a different type (e.g. int8\_t). The developer may have to manually rewrite many HDC operations to operate on the appropriate types as well as fix allocations of data with the appropriate allocation sizes. An example of such a transformation is to set the element type of some hypervectors to a single bit. This may be beneficial when using the \textit{hdc\_sign} HD-primitive in HDC++, which maps elements in hypervectors and hypermatrices to a binary representation in \{+1,-1\}. 

To reduce the burden on the programmer in performing such rewrites, we implement an automatic binarization inter-procedural analysis and transformation described in Algorithm ~\ref{alg:autobin}. The algorithm performs a taint-analysis to identify which HDC operations and allocations are tainted by \textit{hdc\_sign} and then subsequently rewrites them to operate on the reduced element-bitwidth. The reduced bitwidth representation is a parameter of this transformation. For a single bit-representation, the lowering of HDC primitives are handled using bitvector logical operations. Any HDC primitives requiring additional allocation updates are binarized accordingly. By default, automatic binarization only binarizes the results of reducing operations such as \textit{matmul}, \textit{cossim\_similarity}, and \textit{hamming\_distance}, and binarizes both the inputs and outputs of element-wise HDC operations. More aggressive binarization can be toggled by specifying another command-line option which reduces precision for the inputs of reducing HDC operations.

\begin{algorithm}
\caption{Algorithm for Automatic Binarization}\label{alg:autobin}
\begin{algorithmic}[1]
\Procedure{HDCBinarize}{$P$ : \textbf{Program}, BinarizedTy: \textbf{Type}, BinarizeReduce? : \textbf{Bool}}
\State Worklist  $\gets$ GetHDCSignInsts($P$)
\State \textit{// Set of HD operations which directly or indirectly}
\State \textit{// interact with a binarized hypervector or hypermatrix}
\State TaintedOpSet $\gets \{\}$ 
\While{WorkList$\not=\{\}$}%\Comment{We have the answer if r is 0}
\State $I\gets$ Extract(WorkList)
\State TaintedOpSet $\gets$ TaintedOpSet $\bigcup$ $\{I\}$
\If{BinarizeReduce?$\And$IsHDCReduceOp($I$)}
    \State \textit{// Taint both input and output vectors/matrices}
    \State Worklist $\gets$ TaintInputOutput($I$,WorkList)
\ElsIf {IsHDCReduceOp($I$)}
    \State \textit{// Taint only output hypervectors/matrices}
    \State Worklist $\gets$ TaintOutput($I$,WorkList)
\Else
    \State Worklist $\gets$ TaintInputOutput($I$,WorkList)
\EndIf
\EndWhile
\For{$I\in$ TaintedOpSet}
    \hspace{1em}\State \textit{// Replace I with Binarized Primitive}\\
    \hspace{3em}$P \gets$ Binarize($I$, BinarizedTy, BinarizeReduce?) 
    %\hspace{1em}\State \textit{// Binarize any associated allocations}\\
    %\hspace{2em}$P\gets$BinarizeAlloc($I$, BinarizedTy, BinarizeReduce?)
\EndFor
%\State \textbf{return} $P$
\EndProcedure
\end{algorithmic}
\end{algorithm}

\textit{\textbf{Reduction Perforation}}.
Reduction operators are common primitives in HDC algorithms---matrix multiplication is used in random projection encoding, and hamming distance and cosine similarity are used in inference algorithms that identify which representative hypervector a query hypervector is most similar to. Due to the error resilience of HDC, it may be sufficient to perform these operations approximately by skipping elements along the reduction axis. We provide application developers a mechanism for controlling such approximations by exposing an HDC reduction perforation directive, \textit{red\_perf}. The reduction computation may be approximated by performing either a segmented reduction or a strided reduction \cite{loopperforation}. In the former, a contiguous sub-range of elements in hypervectors are used to calculate the reduction, whereas in the latter a strided access of elements in hypervectors are used. The parameters for these approximations are exposed to the application developer in their HDC++ source code. 

Reduction perforation is specified with the \textit{red\_perf} HDC++ primitive. The first argument is the value computed by the reduction primitive the programmer wants to approximate, the second argument is the starting offset into the hypervectors, the third argument is the ending offset into the hypervectors, and the fourth argument is the stride to sample elements at between this range. \textit{red\_perf} acts as a compiler directive, indicating that loops generated for another operation should be modified---it can be applied to \textit{hamming\_distance}, \textit{cossim\_similarity}, \textit{matmul}, and \textit{l2norm}. For \textit{hamming\_distance} and \textit{cossim\_similarity}, only the relative magnitude between (dis)similarities matters in every HDC application we know of. Thus, we don't scale the final (dis)similarity to accommodate for skipping elements. For \textit{matmul} and \textit{l2norm}, we divide accumulated values by the proportion of elements in the original hypervectors that were visited. For these intrinsics, scaling is necessary since their usage is diverse in HDC applications, so their absolute magnitudes matter. Reduction perforation is a specific application of loop perforation \cite{loopperforation} to HDC operations. Listing~\ref{lst:search-approx} illustrates a strided and segmented Hamming distance calculation.

\begin{lstlisting}[language=C, label={lst:search-approx}, caption={Simplified HDC++ example describing a segmented and/or strided approximated Hamming distance.}]
hypervector<2048> A, B = ...;
// Calculate hamming distance in the sub-range [0, 1024)
auto dst_segmented = __hetero_hdc_hamming_distance(A, B);
__hetero_hdc_red_perf(dst_segmented, 0, 1024, 1);
// Calculate hamming distance for every other element
auto dst_strided = __hetero_hdc_hamming_distance(A, B);
__hetero_hdc_red_perf(dst_strided, 0, 2048, 2);
// Hamming distance for every other element in [0, 1024)
auto dst_both = __hetero_hdc_hamming_distance(A, B);
__hetero_hdc_red_perf(dst_both, 0, 1024, 2);
\end{lstlisting}

These two optimizations are fundamentally approximate. Automatic binarization reduces element bitwidth which reduces data-movement between the host and device and also improves throughput by using hardware efficient bit-wise operations. Reduction perforation approximates reduction operators between hypervectors which improves performance by reducing the total number of computations which need to be performed. These are optional transformations in HPVM-HDC; we study their implementation and empirical effects, but do not tackle automatic selection of approximations. However, they are easily explorable by an application developer; automatic binarization is exposed as a compiler flag and reduction perforation is exposed as an optional HDC++ primitive that doesn't affect algorithm code. They are applicable on the CPU and GPU, since these devices have the flexibility to execute HDC operations with different element types and loop iteration spaces. However, they are not applicable on the HDC accelerators, since these devices do not support these approximations.

\subsection{Back End Code Generation}
\label{sec:hd-backend}
When targeting the CPU, HPVM-HDC translates HDC primitives into HPVM IR subgraphs containing data-level parallelism. These subgraphs seamlessly interoperate with other Hetero-C++ parallel constructs used by the programmer, for example to express encoding. Once the entire application is lowered into HPVM IR, HPVM-HDC uses HPVM's CPU backend to generate an executable.

On the GPU, HPVM-HDC takes advantage of optimized routines written in CUDA, cuBLAS \cite{cublas}, and Thrust \cite{thrust}, which are NVIDIA-specific GPU programming interfaces. cuBLAS contains optimized implementations of matrix multiplication, transposition, and normalization, while Thrust contains optimized reduction routines. When targeting NVIDIA GPUs, HPVM-HDC lowers HDC primitives directly to cuBLAS calls, Thrust calls, or CUDA kernels instead of HPVM IR to take advantage of these optimized GPU routines. Generic parallel code expressed using Hetero-C++ primitives are instead lowered to HPVM-IR and then OpenCL code using HPVM's generic GPU back end. In Section~\ref{sec:eval}, we compile an application, HyperOMS, that utilizes both of these lowering schemes.

HPVM-HDC targets two HDC accelerators: a digital ASIC and a ReRAM accelerator simulator. These devices provide course-grain operations for performing HDC encoding, training, and inference. These operations are impractical to automatically identify in programs expressed with granular HDC primitives. Additionally, we want to minimize data movement costs between the host machine and the devices. HPVM-HDC lowers the \textit{encoding\_loop}, \textit{training\_loop}, and \textit{inference\_loop} primitives to calls to the accelerators' low-level functional interface. Listing~\ref{lst:acc_code} shows an example of lowering \textit{training\_loop} and \textit{inference\_loop}. Crucially, HPVM-HDC can lift redundant data movements outside of loops, which would be much more difficult with non-loop based HDC primitives.

While these accelerators implement particular algorithms for HDC encoding, training, and inference, this is not reflected in the high-level HDC++ primitives. This is because the encoding, training, and inference algorithms that could be used on CPUs or GPUs are diverse and require user specification. This user specification comes in the form of an implementation function, built using more granular HDC primitives (as shown in Listing~\ref{lst:stage}). The implementation function is used when targeting CPUs or GPUs, while high-level HDC operations are leveraged when targeting HDC accelerators. This has two benefits. First, HDC++ applications can achieve competitive performance on all targets, as HPVM-HDC already compiles the more granular HDC primitives in the implementation function to generate efficient code. Second, the rest of the HDC application can be built irrespective of whether encoding, training, or inference are actually performed on a CPU, GPU, or an HDC accelerator.

\begin{lstlisting}[language=C, label={lst:acc_code}, caption={HPVM-HDC generated code for performing HDC training and inference on the digital HDC ASIC.}]
config_t config = ...;
initialize_device(&config);
allocate_base_mem(random_projection);
allocate_class_mem(classes);
for (int n = 0; n < EPOCHS; ++n)
	for (int i = 0; i < N_TRAIN; ++i)
		allocate_feature_mem(train_inputs[i]);
		execute_retrain(train_labels[i]);  
read_class_mem(classes);
allocate_base_mem(random_projection);
allocate_class_mem(classes);
for (int i = 0; i < N_TEST; ++i)
	allocate_feature_mem(infer_inputs[i]);
	infer_labels[i] = execute_inference();
\end{lstlisting}

\section{Evaluation}
\label{sec:eval}

\subsection{Methodology}
\label{sec:eval-methodology}

\textit{\textbf{Hardware targets.}}
The evaluation setup for CPU and GPU consists of an Intel(R) Xeon(R) Silver 4216 CPU with 16 cores, 32 threads, and 192GB of memory and an NVIDIA GeForce RTX 2080 Ti GPU with 11GB of memory. We execute the digital HDC ASIC as an accelerator attached to an ARM core as the host through a custom FPGA-based interface. We execute the ReRAM HDC accelerator simulator on the Intel machine---the simulator calculates an estimate of the latency of a taped out ReRAM accelerator. For comparison with the HDC accelerators, we also execute HDC++ applications on an NVIDIA Jetson AGX Orin 64GB board with an Ampere GPU with 2048 CUDA cores and 64 tensor cores---this comparison is meant to be more representative of hardware available at the edge, which is the target use case of these HDC accelerators (the Jetson AGX Orin is a moderately powerful GPU relative to other edge-class GPUs).

\textit{\textbf{Metrics and measurements.}}
We measure the wall clock time as a mean of 3 repeated executions on an otherwise idle system.

\begin{table*}[t]
  \centering
  \begin{tabular}{|c|>{\centering\arraybackslash}m{0.36\linewidth}|>{\centering\arraybackslash}m{0.36\linewidth}|}
  \hline
    \textbf{HDC Applications} & \textbf{Workload Significance} & \textbf{HDC Stages Used} \\
    \hline
    HD-Classification \cite{HD2FPGA} & Classification implemented using HDC & Random-Projection Encoding, Inference, Training \\
    \hline
    HD-Clustering \cite{HD2FPGA, hdclustering} & K-means clustering implemented using HDC & Random-Projection Encoding, Inference \\
    \hline
    HyperOMS \cite{hyperoms} & Open Modification Search for Mass Spectrometry & Level ID Encoding, Inference \\
    \hline
    RelHD \cite{relhd} & GNN learning, data relationship analysis & Graph Neighbor Encoding, Inference, Training \\
    \hline
    HD-Hashtable \cite{biohd} & Genome Sequence Search for Long Reads Sequencing using HD hashing &  K-mer based Encoding, Inference \\
    \hline
  \end{tabular}
  \caption{Summary of HDC applications implemented in HDC++ and evaluated using HPVM-HDC. Applications using HDC require similar stages, however the specific algorithms used vary across application domains.}
  \label{tab:HDC-workloads}
\end{table*}

\textit{\textbf{Applications:}}
We implement five applications, summarized in Table~\ref{tab:HDC-workloads}, in HDC++. All five applications were developed by domain experts in prior work. HD-Classification and HD-Clustering represent common computational kernels in HDC applications, while HyperOMS, RelHD, and HD-Hashtable are three existing HDC applications from prior work. Our evaluated workloads capture most of the core computations present in other HDC applications, such as HyperSpec \cite{hyperspec}, SpecHD \cite{spechd}, HDBind \cite{hdbind}, HyperGen \cite{HyperGen}, and RapidOMS \cite{rapidoms}. HD-Classification and HD-Clustering codes were developed and tuned as part of \cite{HD2FPGA} and \cite{hdclustering}, respectively. We used the reference implementations available for Rel-HD \cite{relhd} (CPU and GPU) and HyperOMS \cite{hyperoms} (just GPU). The HD-Hashtable code is a modified version of \cite{biohd}, optimized for using hash tables. We evaluate HD-Classification and HD-Clustering on the Isolet \cite{isolet} dataset of spoken letters, HyperOMS on  the Yeast and Human Spectral Library dataset and the iPRG2012 dataset \cite{hyperomsquery,hyperomsref}, RelHD on the Cora dataset \cite{cora} of categorized scientific papers, and HD-Hashtable on Genomics dataset adapted from \cite{Bankevich2022-vs}. When targeting the GPU, these baseline codes are often written in CUDA C++. While unportable, this enables the use of specific device features that improve performance, such as tensor cores and warp voting functions.

\subsection{Performance}
\label{sec:eval-perf}

To evaluate HPVM-HDC's performance across architectures, we compile and execute the aforementioned applications on the four hardware targets listed above. We evaluate HD-Classification and HD-Clustering on all 4 hardware targets, and HyperOMS and RelHD on the CPU and GPU only. The HDC accelerators expose coarse-grain computations for random projection encoding, training, and inference. HD-Classification maps efficiently to all three of these operations, while HD-Clustering maps to random projection encoding and inference. The other three applications do not map to these particular coarse-grained operations; we are therefore unable to execute those three applications on the HDC custom accelerators. 

\begin{figure}
    \centering
\includegraphics[width=0.46\textwidth]{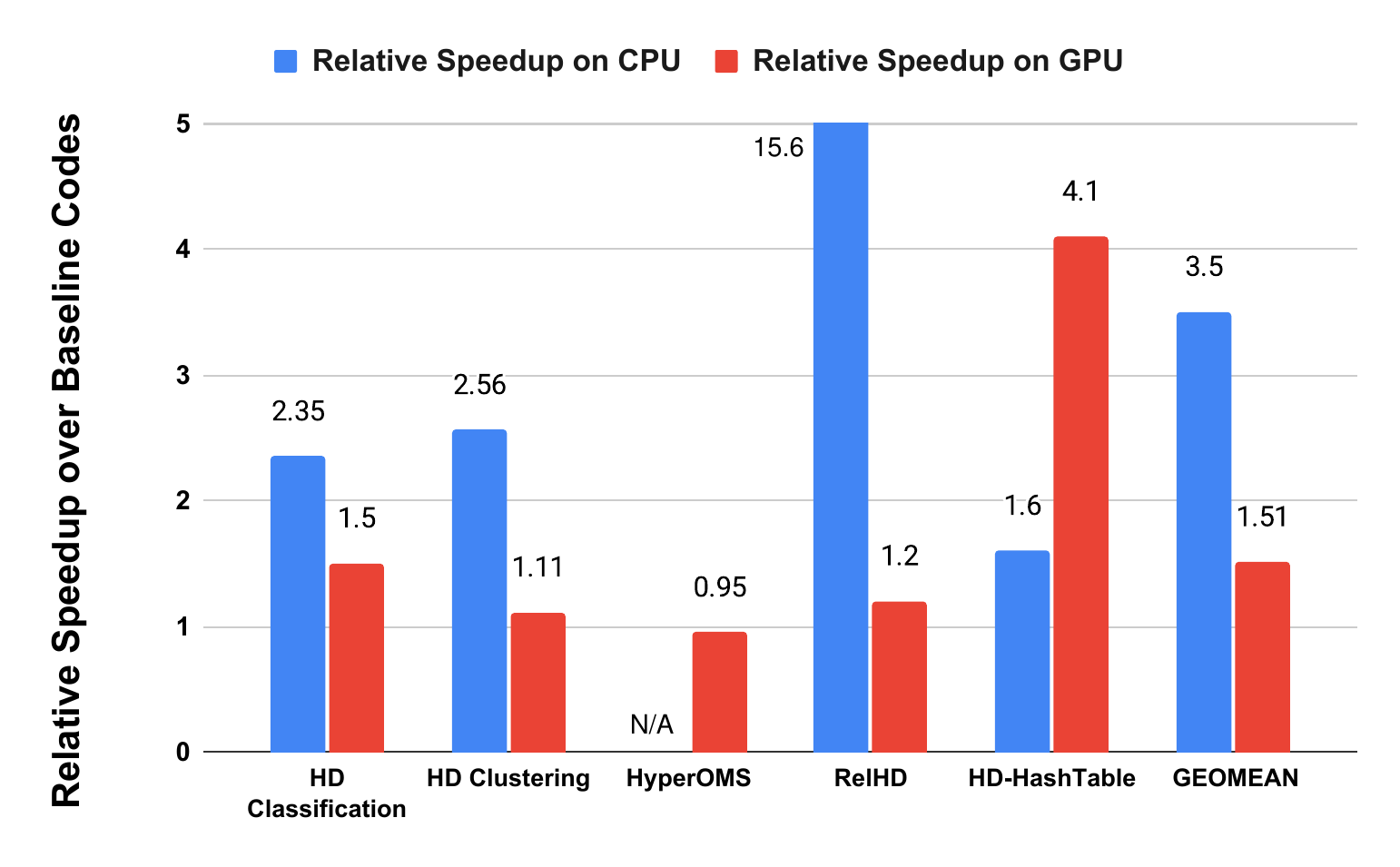}
%\vspace{-.1in}
    \caption{\small{Performance of applications written in HDC++ and compiled with HPVM-HDC for execution on CPU and GPU, compared against hand-written device-specific baseline codes. The relative speed-up of the HDC++ implementation is shown, higher is better. }}
    \label{fig:performance}
\end{figure}

Figure~\ref{fig:performance} shows the relative speedup in wall clock time for each application on CPU and GPU when compiled with HPVM-HDC, compared to the baseline device-specific codes for these devices from prior work. HyperOMS lacks a CPU version we can use as the baseline---we omit the relative speedup bar in that case.  In contrast, all of the applications can be compiled for and run on CPUs and GPUs using HPVM-HDC. HPVM-HDC achieves significant speed-ups over several baseline codes across architectures without using approximation-based optimizations. However, this advantage is a function of the specific implementation and tuning choices of those baselines and not a fundamental advantage of HPVM-HDC. The key contribution of HPVM-HDC is achieving \textit{competitive} performance across applications and architectures while presenting a high-level, target-agnostic programming interface in HDC++.

\textit{\textbf{Performance on CPU.}}
HPVM-HDC targets CPUs by lowering HPVM IR into sequential machine code. Each application with a CPU baseline uses Python and NumPy. We do not draw conclusions with regards to performance improvements on the CPU as the reference implementations are interpreted in Python while HPVM-HDC codes are compiled ahead-of-time, which inherently does not suffer from the overheads of the Python interpreter. %We do include the relative performance on CPU in Figure~\ref{fig:performance} to demonstrate that HPVM-HDC is able to compile HDC++ applications onto CPUs.
%HPVM-HDC achieves a geomean speedup of 3.5x compared to these baselines, since it generates native machine code implementing HDC primitives and does not suffer from the overheads of the Python interpreter.

\textit{\textbf{Performance on GPU.}}
For GPUs, HPVM-HDC lowers HDC primitives into calls to CUDA kernels, cuBLAS functions, or Thrust functions, and lowers generic parallel HPVM IR to OpenCL using HPVM's GPU backend. The only application that is slower when written in HDC++ is HyperOMS, which is 5\% slower than the baseline.  The bottleneck portion of HyperOMS in our experiments is level ID encoding. In our HDC++ implementation, this is written using generic Hetero-C++ parallel constructs which generate OpenCL using HPVM's GPU backend. In contrast, the baseline implementation uses an optimized CUDA kernel, including the use of warp-level primitives that HPVM cannot generate.

Similar to the CPU compiled benchmarks, HD-Hashtable's GPU compilation is interpreted in Python using CuPy, whereas HPVM-HDC generates a compiled executable. HD-Classification, HD-Clus\-tering, HyperOMS and RelHD provide a CUDA C++ baseline implementation, so the comparisons for these can be used to derive meaningful performance insights. Across the GPU compiled HPVM-HDC applications for which the reference implementation is in CUDA C++, we observe a range of 0.95x -- 1.5x.  
The largest performance improvement (1.5x) is in HD-Classification. For these 4 benchmarks, HPVM-HDC achieves a geomean 1.17x relative speedup. This speedup is primarily the result of different tuning choices in HPVM-HDC generated code compared to the baseline codes.

%HD-HashTable is the application with the largest performance difference between the baseline and HDC++ implementations. HD-HashTable is the only application whose baseline is written in Python and uses CuPy to target the GPU. All the other baseline GPU implementations directly use CUDA, cuBLAS, and Thrust, and thus don't suffer from the runtime overhead of the Python interpreter.

\begin{figure}
    \centering
\includegraphics[width=0.46\textwidth]{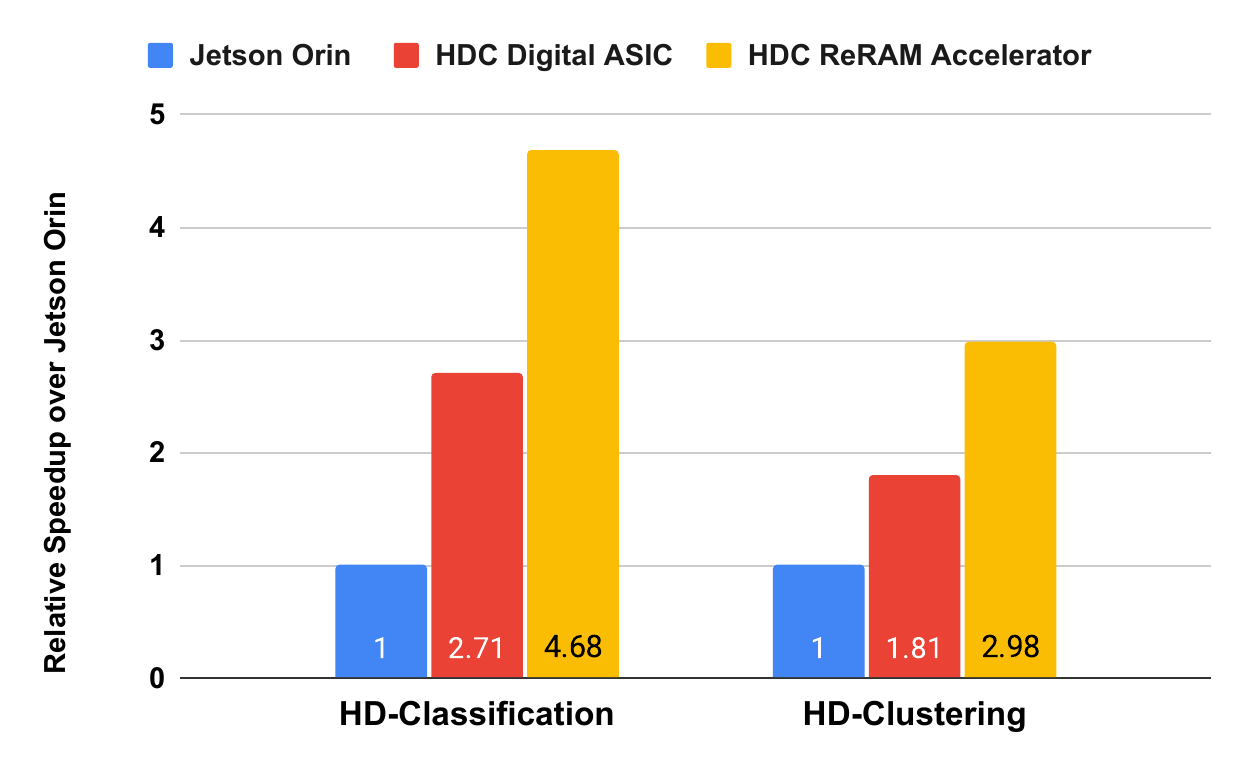}
%\vspace{-.1in}
    \caption{\small{Performance of applications written in HDC++ and compiled with HPVM-HDC for execution on the HDC Digital ASIC and the HDC ReRAM Accelerator simulator, compared against the same code executed on an NVIDIA Jetson Orin board, higher is better.}}
    \label{fig:acc-performance}
\end{figure}

\textit{\textbf{Performance on HDC Accelerators.}}
Figure~\ref{fig:acc-performance} compares the performance of HD-Classification and HD-Clustering when compiling HDC++ applications for the HDC accelerators and for an NVIDIA Jetson AGX Orin board. Due to fabrication cost constraints, the digital ASIC and its ARM host CPU only communicate at approximately 10 kbps. Additionally, the ReRAM accelerator simulator only models the performance of the ReRAM accelerator, and not of a system containing the accelerator and a host. Thus, we measure the "\textit{device-only}" performance for the accelerators and the Jetson Orin, which corresponds to just the HDC primitive code in the applications.

We compare running HD-Classification and HD-Clustering on HDC accelerators against running on an NVIDIA Jetson AGX Orin board, since there is no baseline implementation targeting the accelerators for comparison---no prior work has run a whole HDC application on either of these accelerators. Additionally, while these accelerators were designed for classification ~\cite{DigitalASIC,ReRAMAcc}, we can accelerate the computationally intensive portions of HD-Clustering by using HDC++ primitives. The NVIDIA Jetson AGX Orin board is representative of GPU-based compute available in edge environments---the HDC accelerators are also designed for use at the edge. We see that the accelerators deliver speed-ups for both HD-Classification and HD-Clustering compared against execution on a GPU-based system. The larger speedup in the case of HD-Classification is because this application spends most of its time training a set of representative hypervectors per class, which the HDC devices are specifically designed to accelerate. The ReRAM accelerator is estimated to be faster than the digital ASIC because the underlying processing-in-memory technology, resistive RAM, provides high density on-chip storage with fast read speeds.

\begin{table*}[t]
\small
  \centering
  \begin{tabular}{|c|c|c|p{7cm}|}
  \hline
    \textbf{ID} & \textbf{Optimization Setting} & \textbf{Required LOC changes} & \textbf{Description} \\
    \hline
    I & Cosine Similarity (Baseline) & 0 & Inference using 32-bit floats with cosine similarity \\
    \hline
    II & Hamming Distance  & 1 & Inference using 32-bit floats with Hamming Distance \\
    \hline
    III & Auto Binarize (Enc + Out)& 1 & Binarization of Class \& Encoded HVs with Hamming Distance \\
    \hline
    IV & Auto Binarize (Enc + In/Out) & 1 & III with casting input features to 32-bit Ints before encoding \\
    \hline
    V & Auto Binarize (Enc + Out + Strided Matmul [2]) & 2 &III with loop perforated matrix multiplication with stride of 2\\
    \hline
    VI & Auto Binarize (Enc + Out + Strided Matmul [4]) & 2 & III with loop perforated matrix multiplication with stride of 4\\
    \hline
    VII & Auto Binarize (Enc + Out + Strided Hamming [2]) & 3 & III with loop perforated Hamming Distance with stride of 2\\
    \hline
    VIII & Auto Binarize (Enc + Out + First Half Hamming) & 3 &III with Hamming Distance only on first half of  hypervectors\\
    \hline
    IX & Cosine Similarity (Strided Encoding [2]) & 1 & I with encoding loop perforated with stride 2 \\
    \hline
    X & Cosine Similarity (Strided Similarity [2]) & 1 & I with cosine similarity loop perforated with stride 2 \\
    \hline
  \end{tabular}
  \caption{\small{Descriptions of optimization settings compiled using HPVM-HDC for HD-Classification Inference.}}
  \label{tab:approx-legend}
\end{table*}

\begin{table*}[t]
\small
    \centering
    \begin{tabular}{|c|c|c|c|c|c|}
       \hline
       \multirow{2}{*}{\textbf{Application}} & \multicolumn{2}{c|}{\textbf{CPU}} & \multicolumn{2}{c|}{\textbf{GPU}} & \\
       \cline{2-6}
       & \textbf{Baseline Lang.} & \textbf{ LOC} & \textbf{Baseline Lang.} & \textbf{LOC} & \textbf{HDC++ LOC} \\
       \hline
       HD-Classification  & Python & 193 (0.47x) & CUDA & 608 (1.48x) & 410 \\
       \hline
       HD-Clustering & Python & 426 (1.24x) & CUDA & 429 (1.25x) & 343 \\
       \hline
       HyperOMS & \multicolumn{2}{c|}{N/A} & CUDA & 1188 (2.12x) & 560 \\
       \hline
       RelHD & Python & 740 (1.15x) & CUDA & 457 (0.71x) & 642 \\\hline
       HD-Hashtable & Python & 164 (0.58x) & Python (CuPy) & 164 (0.58x) & 283 \\
       \hline

    \end{tabular}
    \caption{\small{Lines of code to express each application for each target device. The CPU and GPU counts correspond to separate baseline programs (except for HD-Hashtable, where both versions use the same Python code), while the HDC++ count corresponds to a single implementation targeting CPUs, GPUs, and HDC accelerators. Reduction factors between the baselines and HDC++ are shown in parentheses, higher is better.}}
    \label{tab:loc}
\end{table*}

\subsection{Impact of HDC optimizations}
\label{sec:eval-opt}
To evaluate the effectiveness of domain specific optimizations in HPVM-HDC, we compile the inference stage of HD-Classification with different optimization settings (Table \ref{tab:approx-legend}). Note that these optimizations are designed to exploit the error resilience in the HDC paradigm, and thus affect not only performance but also classification accuracy. HD-Classification in particular uses label data to assess the accuracy of a classifier, so it is straight-forward to measure the end-to-end effect of approximating inference. Through the HDC++ programming model and parameterized HDC transformations in HPVM-HDC, application developers can rapidly explore different approximation configurations in their applications with minimal source code changes (1-2 lines of code). 

\begin{figure}
    \centering
\includegraphics[width=0.46\textwidth]{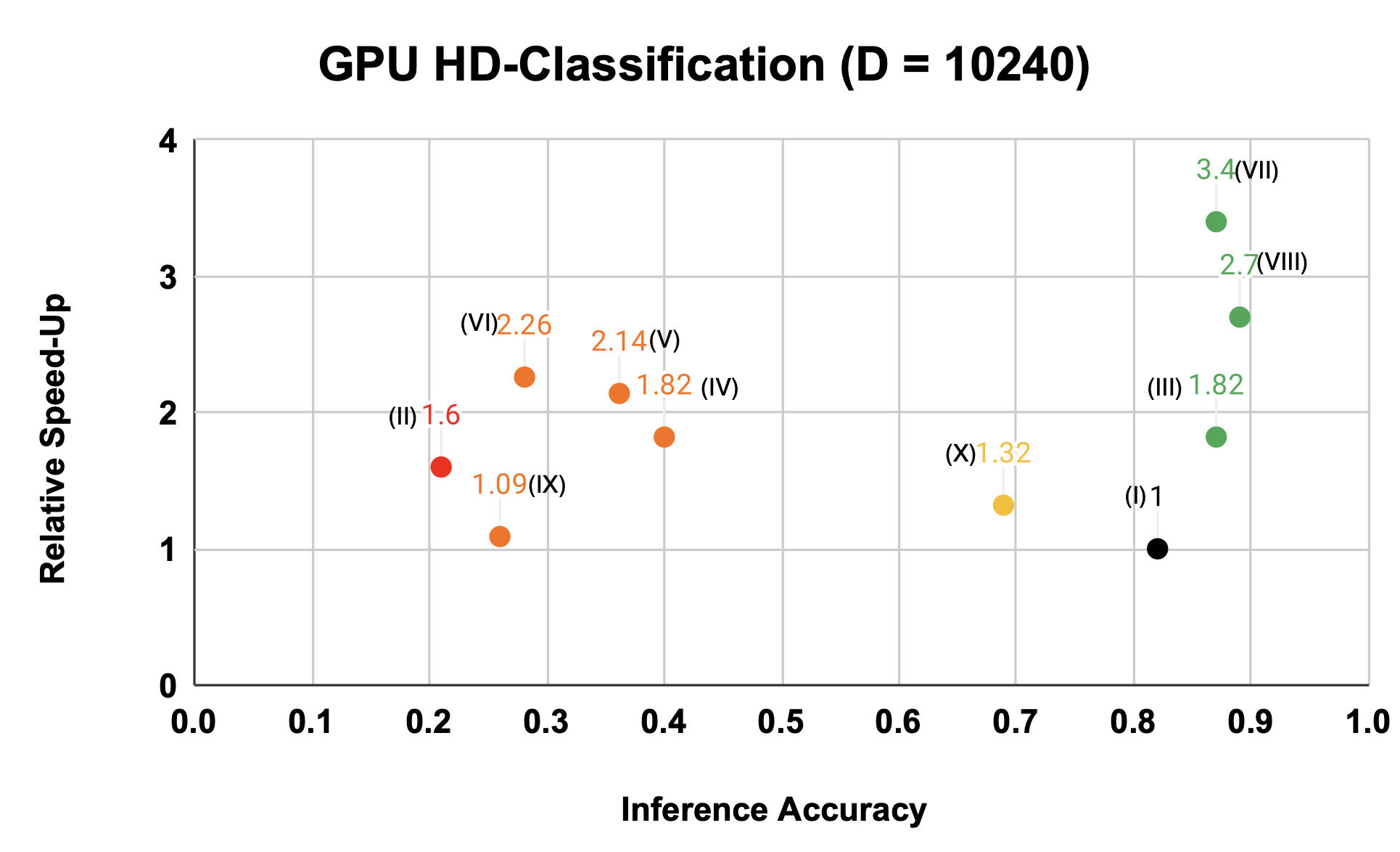}
    \caption{
    \small{Speedup versus accuracy for HD-Classification inference on the GPU. The black point represents the baseline configuration. All approximation configurations achieve improved performance. Green points denote configurations maintaining high accuracy. Yellow points denote configurations with moderate accuracy loss and red points denote configurations with significant accuracy loss.}
    }
    \label{fig:approx-gpu}
\end{figure}

For this evaluation, we set the encoding dimension to 10240 and derived the trained class hypervectors using cosine similarity in a single pass over the training dataset offline. Figure~\ref{fig:approx-gpu} plots the relative speed-up versus inference accuracy for different optimization settings described in Table~\ref{tab:approx-legend}. These configurations are not meant to be exhaustive but merely illustrative of points in the various stages in an HDC application which can be approximated. The baseline \textbf{(I)} achieves an inference accuracy of 0.82. While all optimization settings achieve a significant performance improvement, the inference accuracy varies drastically ranging from a low of 0.2 to a high of 0.89. The points in green (\textbf{III, VII, VIII}) result in improved accuracy over the baseline, as well as faster performance. When using the binarization approximation, i.e., projecting the floating point values in the class hypervectors to \{+1, -1\}, which are represented using 1-bit elements, and using Hamming distance similarity measure, we observe a negligible change in accuracy. This aligns with prior observations regarding the effects of binarized learning on accuracy \cite{binarylearning}. Additionally, this reduces the initial data movement necessary as the class hypervectors are represented using fewer bits in total. Reduction perforation adds upon the performance improvements achieved from automatic binarization. However, the downstream effect on accuracy varies significantly depending on which operation is being perforated. Configurations (\textbf{VII, VIII, X}) perforate the inference's similarity computation and result in at most a 13\% drop in accuracy with respect to the baseline in \textbf{X} and no drop in the other two. However, when perforating the encoding stages (\textbf{V, VI, IX}), the inference accuracy drops to 25-35\%. This implies that for HD classification, the similarity computation is much more amenable to approximation than the encoding stage.

\subsection{Programmability}
\label{sec:eval-programmability}
\textit{\textbf{Source code portability}}: HPVM-HDC can compile a single HDC++ source program to multiple hardware targets, \textit{with no target-specific tuning or customization}. This is a valuable benefit because adoption of custom hardware accelerators for real-world applications will be far more practical if the applications do not have to be rewritten for each hardware target: in fact, we believe this is a crucial requirement for enabling success for custom hardware, as shown by the widespread success of custom accelerators for machine learning which has been strongly enabled by hardware-independent programming frameworks like PyTorch \cite{pytorch} and TensorFlow \cite{tensorflow}.

\textit{\textbf{Lines of Code}}: The high-level HDC operations in HDC++ also enable more concise code.  We compare the lines of code required to implement the applications with their respective baselines. Note that for each target architecture, there is a different baseline version, often implemented in a different programming language. Table ~\ref{tab:loc} describes the total lines of code reduction in HDC++ versus the baselines for the applications. HDC++ provides a 1.6x geomean reduction in total lines of code (by combining the lines of codes across all baseline target implementations) for HDC applications while remaining retargetable. The HD-Classification and HD-Hashtable baseline CPU codes, which are implemented in Python, are smaller than the HDC++ application codes. This is a consequence of HDC++ being embedded in C++ which is inherently more verbose than Python. However, the HD-Classification, HD-Clustering, and HyperOMS baseline GPU codes, which are implemented in CUDA C++, are larger than the HDC++ application codes; HDC++ abstracts away data-movement between the device and host memory. Additionally, the high-level HDC++ primitives compactly represent the corresponding manually written kernel functions in the GPU baselines, resulting in fewer lines of code.

\textit{\textbf{Approximation Configuration}}: The HDC++ programming interface and HPVM-HDC compiler make it easy to fine-tune HDC applications to balance performance and accuracy because the approximation configuration can be separately specified with almost no changes to the application source code. This includes experimenting with different similarity metrics, encoding strategies, and transformations implemented in HPVM-HDC evaluated in Section ~\ref{sec:eval-opt}. 

To evaluate the improved programmability of approximations in HDC++, we \emph{manually} implemented three of the automatic transformations described in Table~\ref{tab:approx-legend} in the GPU baseline code for HD-Classification. For configuration \textbf{X}, only 4 lines of code needed to be modified. Implementing \textbf{II} required 25 lines of code and approximately 30 minutes of work. Finally, implementing \textbf{III} required the most source code modifications---this configuration requires modifying the allocations of the class hypermatrix and encoded hypervectors, implementing packing elements to 1-bit, and implementating Hamming distance on 1-bit elements. 33 lines of code were written requiring approximately 1 hour. In contrast, HPVM-HDC enables exploring these transformations within seconds by modifying 1-2 lines of the application code. Furthermore, these transformations occur in the HPVM-HDC IR representation, allowing for CPU and GPU implementations of the approximations to be automatically generated, whereas the manual modifications are specific to the GPU baseline.

\section{Related Work}
\label{sec:relwork}

\textit{\textbf{Languages for HDC.}} Because Python \cite{van1995python} and MATLAB \cite{matlab2012matlab} enable fast prototyping, these languages are used to implement the majority of experimental research HDC algorithms and libraries~\cite{verges2023hdcc}. Some HDC algorithms are also implemented in C++. HDC algorithms are expressed in these languages as sequential scalar code or using array processing libraries such as NumPy. These languages can be used to target GPUs using language extensions \cite{tuomanen2018hands, fatica2007accelerating, nozal2021exploiting}. HDC algorithms are also written in CUDA \cite{sanders2010cuda} and OpenCL \cite{munshi2009opencl} for GPUs \cite{jin2019accelerating, kim2020geniehd}, and using C++ with FPGA-specific extensions to target FPGAs \cite{salamat2019f5, imani2019fach}. However, no HDC-specific heterogeneous language or extensions to existing languages exist yet in the literature. 

\textit{\textbf{Libraries for HDC.}} Work on libraries for HDC is relatively extensive. TorchHD \cite{heddes2022torchhd} is a library implemented using PyTorch \cite{paszke2019pytorch} and provides interfaces for implementing HDC algorithms on CPUs and GPUs. OpenHD \cite{kang2022openhd} is a library implemented in Python with CUDA extensions to facilitate implementation of HD Classification and Clustering for GPUs. HDTorch \cite{simon2022hdtorch} is another library implemented in Python with CUDA extensions for hypervector operations for classical and online HDC learning. None of these libraries support, nor can be easily extended to target, HDC accelerators.

\textit{\textbf{Compilers for HDC.}} The HDCC compiler \cite{verges2023hdcc} is the first and, to date, the only compiler we know of developed for HDC. However, HDCC supports a very limited programming model. HDCC supports a fixed and limited set of encoding, training, and inference algorithms, rather than a set of lower level hypervector and hypermatrix primitives like HPVM-HDC. Another limitation of HDCC is that it does not support heterogeneous compilation: it only supports one back end that generates C code with POSIX \cite{walli1995posix} and vector extensions for multi-core CPUs from high-level descriptions of HDC algorithms.  Unlike HPVM-HDC, it does not target other hardware architectures, which present more lucrative performance and energy efficiency trade-offs.

\textit{\textbf{Compilers for Deep Learning.}} Work on compilers for deep learning is extensive \cite{xla, tvm, glow, dlcompiler}. These compilers operate on high-level descriptions of deep learning models, often defined in high-level frameworks such as Tensorflow \cite{tensorflow} and PyTorch \cite{pytorch}. Programming HDC algorithms requires tweaking and exploring different algorithmic and approximation choices which can only be facilitated by allowing programmers to express custom algorithms in addition to providing pre-ordained high-level HDC abstractions. This flexibility, however, is not afforded by deep learning frameworks. Moreover, deep learning models are compiled to computational dataflow graphs in XLA \cite{xlaops}; this representation does not adequately capture all forms of parallelism, unlike HPVM-HDC IR which supports a hierarchical dataflow representation that captures multiple kinds of parallelism.

%These frameworks typically expose tensor operators as primitives--expressing control flow around these tensor operators and exploiting parallelism in that control flow can be challenging. In contrast, HDC++ is embedded in Hetero-C++, which allows the user to express general parallel code in their applications while using primitives that describe matrix and vector operations. This is particularly critical for the encoding stage of many HDC applications, such as HyperOMS, where having support for parallel outer loops is critical. Additionally, the algorithm stage primitives exposed in HDC++ are higher level than individual operators that are traditionally operated on in deep learning compilers. 

\section{Conclusion}

This paper presents the first heterogeneous programming system for HDC. We introduce a novel programming language, HDC++, for writing HDC applications using a unified programming model with HDC-specific primitives. We also introduce a heterogeneous compiler, HPVM-HDC, that provides a retargetable intermediate representation for compiling HDC programs to many hardware targets. We demonstrate the implementation and evaluate the benefits of two HDC-specific software optimizations that balance end-to-end quality of service and performance. HPVM-HDC achieves better performance than baseline HDC applications on CPUs and GPUs with less engineering effort. Our programming system enables for the first time running HDC applications on two HDC accelerators. HDC++ and HPVM-HDC enable simultaneous and decoupled development of HDC applications and accelerators, allowing for faster development and adoption of both. We anticipate that HPVM-HDC will be straightforward to extend for future developments in HDC accelerators---for example, some proposed devices support hardware-based tuning parameters \cite{SpecPCM} that HPVM-HDC could support in addition to its software-based optimizations.

\begin{acks}
We thank the anonymous reviewers for their helpful feedback on this paper. This work was supported by funding from PRISM, one of the seven centers in JUMP 2.0, a Semiconductor Research Corporation (SRC) program sponsored by DARPA.
\end{acks}

\bibliographystyle{ACM-Reference-Format}
\bibliography{refs}
\end{document}